\documentclass[conference]{IEEEtran}
\IEEEoverridecommandlockouts
\usepackage{cite}
\usepackage{amsmath,amssymb,amsfonts}
\usepackage{algorithmic}
\usepackage{graphicx}
\usepackage{textcomp}
\usepackage{xcolor}
\usepackage{booktabs}
\usepackage{dirtree}
\usepackage{tikz}
\def\BibTeX{{\rm B\kern-.05em{\sc i\kern-.025em b}\kern-.08em
    T\kern-.1667em\lower.7ex\hbox{E}\kern-.125emX}}

    \newcommand\copyrighttext{%
  \footnotesize \textcopyright 2024 IEEE. Personal use of this material is permitted.
  Permission from IEEE must be obtained for all other uses, in any current or future
  media, including reprinting/republishing this material for advertising or promotional
  purposes, creating new collective works, for resale or redistribution to servers or
  lists, or reuse of any copyrighted component of this work in other works.}
\newcommand\copyrightnotice{%
\begin{tikzpicture}[remember picture,overlay]
\node[anchor=south,yshift=10pt] at (current page.south) {\fbox{\parbox{\dimexpr\textwidth-\fboxsep-\fboxrule\relax}{\copyrighttext}}};
\end{tikzpicture}%
}

\begin{document}

\title{Vulnerable Road User Clustering for Collective Perception Messages: Efficient Representation Through Geometric Shapes}


\author{\IEEEauthorblockN{Edmir Xhoxhi}
\IEEEauthorblockA{
\textit{Institute of}\\
\textit{Communications} \\
\textit{Technology}\\
\textit{Leibniz University}\\
Hannover, Germany \\
edmir.xhoxhi@\\
ikt.uni-hannover.de}
\and
\IEEEauthorblockN{Vincent Albert Wolff}
\IEEEauthorblockA{
\textit{Institute of}\\
\textit{Communications} \\
\textit{Technology}\\
\textit{Leibniz University}\\
Hannover, Germany \\
vincent.wolff@\\
ikt.uni-hannover.de}
\and
\IEEEauthorblockN{Yao Li}
\IEEEauthorblockA{
\textit{Institute of}\\
\textit{Cartography}\\
\textit{and Geoinformatics}\\
\textit{Leibniz University}\\
Hannover, Germany \\
yao.li@ \\
ikg.uni-hannover.de}
\and
\IEEEauthorblockN{Florian Alexander Schiegg}
\IEEEauthorblockA{
\textit{Digital Mobile}\\
\textit{Communication}\\
\textit{and V2X Systems} \\
\textit{Robert Bosch GmbH}\\
Hildesheim, Germany\\
florian.schiegg@\\
de.bosch.com}
}

\maketitle
\copyrightnotice


\begin{abstract}
Ensuring the safety of Vulnerable Road Users (VRUs) is a critical concern in transportation, demanding significant attention from researchers and engineers. 
Recent advancements in Vehicle-to-Everything (V2X) technology offer promising solutions to enhance VRU safety. 
Notably, VRUs often travel in groups, exhibiting similar movement patterns that facilitate the formation of clusters. 
The standardized Collective Perception Message (CPM) and VRU Awareness Message in ETSI's Release 2 consider this clustering behavior, allowing for the description of VRU clusters. 
Given the constraints of narrow channel bandwidth, the selection of an appropriate geometric shape for representing a VRU cluster becomes crucial for efficient data transmission.
In our study we conduct a comprehensive evaluation of different geometric shapes used to describe VRU clusters. 
We introduce two metrics: Cluster Accuracy (CA) and Comprehensive Area Density Information (CADI), to assess the precision and efficiency of each shape.
Beyond comparing predefined shapes, we propose an adaptive algorithm that selects the preferred shape for cluster description, prioritizing accuracy while maintaining a high level of efficiency. 
The study culminates by demonstrating the benefits of clustering on data transmission rates. 
We simulate VRU movement using real-world data and the transmission of CPMs by a roadside unit. 
The results reveal that broadcasting cluster information, as opposed to individual object data, can reduce the data transmission volume by two-thirds on average. This finding underscores the potential of clustering in V2X communications to enhance VRU safety while optimizing network resources.
\end{abstract}


\section{Introduction}
\label{sec:intro}

The critical need to protect Vulnerable Road Users (VRUs) has driven engineers and researchers to explore innovative safety enhancements.
Emerging Vehicle-to-Everything (V2X) technologies, such as the Collective Perception Service (CPS) and the VRU Awareness Service (VAS), are pivotal in enhancing VRU safety by improving overall awareness.
The CPS leverages vehicle-mounted sensors to disseminate data on traffic entities and environmental conditions, prioritizing VRUs for inclusion in Collective Perception Messages (CPM).
This information is than used by the receiving vehicles in order to enrich their local environment.


An notable characteristic of VRUs is their tendency to move in clusters or groups.
This characteristic offers a significant opportunity to enhance safety measures if utilized effectively.
Given the critical importance of VRU safety, efficiently communicating the presence of these groups can not only facilitate safety applications but also optimize the use of communication channels.
In Release 2, the VAM ETSI standard \cite{2etsiVAM} has added clustering features that capitalize on the VRUs' behavior by incorporating a cluster information container and a cluster operation container.
With these containers, the cluster leader, once received information from other VRUs can aggregate and broadcast this information about the entire group.
Similarly, CPM's newest release \cite{2etsiCPM} includes a perceived object container.
This container allows the sharing of data on objects detected by onboard sensors through network connectivity.
These objects span various categories, including passenger or service vehicles, VRUs, and even entire VRU clusters.
Furthermore, CPMs can be disseminated by RSUs, expanding the range of information sharing and enhancing the safety of VRUs on the road.

In this work, we aim to assess the significance of chosen geometric shapes utilized to represent clusters of VRUs.
Our analysis will reveal the influence that the usage of specific shapes has on both the accuracy and efficiency of cluster description.
Additionally, we will introduce an adaptive shape algorithm designed to select the most efficient shape based among those with the highest accuracy in representing VRU clusters.
Finally, our study adopts a more holistic approach to illustrate the benefits of clustering for reducing channel load in communication networks.
The empirical foundation of our work is based in a practical scenario where a RSU which transmits CPMs is deployed along pedestrian pathways.
The positions and movements of pedestrians are derived from a real-world dataset.
This setup allows us to closely simulate real-world conditions and evaluate the effectiveness of clustering in communication efficiency for VRUs.

The rest of the paper is organized as follows: in the next section we lay out the related work done in the field.
In Section \ref{sec:scenarioOutline} we describe the scenario outline, while describing the dataset and the clustering algorithm used in this work.
In Section \ref{sec:clusterShapes} a thorough evaluation on shapes used for the cluster description has been carried out.
In Section \ref{sec:results} a more holistic view on the scenario has been taken, were we show the positive impact that VRU clustering has on the amount of sent data.
In the last section we summarize the results and give a short overview of the future outlook.

\section{Related Work}
\label{sec:BackgroundRelWork}

In the context of V2X and clustering of entities for efficient information aggregation, two different approaches exist. The formation of a cluster can either be done decentrally by groups of VRUs, or done by an independent transportation station systems, such as road side units, forming the cluster from a bird-eye view perspective.
Due to the recent advances in standardization of the VAM, first research results were presented for decentral approaches. Rupp et al. \cite{rupp2023} conducted a simulation study on self-organized clustering of Vulnerable Road Users in VANets.
They investigated the \textit{maxClusterVelocityDifference} parameter of VAMs, concluding that this threshold may need to be increased in the VAS standard for effective clustering. Furthermore clustering leads to increased position error, causing a trade-off between number of sent VAMs and position accuracy. Similarly, Lobo et al. \cite{lobo2023}  simulated VAM clustering and measure packet error rate, position error and the impact of Decentralized Congestion Control (DCC) on clustering.
In order to reduce complexity, the buffer shape was pre-set to be circular.
The authors proposed an additional shape buffer to include VRU dynamics, based on the VAM transmission time and VRU velocity.

According to our finding, central approaches, directly addressing the clustering problem in context of V2X, remain an open research topic. To leverage the benefits of a centralized information unit as represented by an RSU, there are, however, alternative approaches. A so-called occupancy grid discredits the positions where vehicles and VRUs can be located, in order to transmit CPMs of smaller size at high traffic participant density. Merwaday et al. show a demo of such an approach \cite{merwaday21}.

In the domain of Collective Perception, research primarily centered on enhancing awareness and safety for Vulnerable Road Users (VRUs) remains relatively underexplored. Willecke et al. \cite{willecke2021vulnerable} performed a simulation study with emphasis on VRU awareness in the CPS. They concluded that without much additional channel resources, the awareness of vehicles in the context of VRUs can be enhanced significantly. Feifel et al. \cite{feifel2023reducing} investigate potential reduction of road crash fatalities, with a focus on non-line-of-sight scenarios involving vehicles and VRUs. The study emphasizes the role of the CAS, also refered to as basic V2X, and CPS in improving road safety. CAS improves situational awareness between equipped vehicles, but its impact is limited by slow adoption rates and challenges in VRU participation. Collective Perception, on the other hand, extends protection by sharing sensor data among vehicles and infrastructure, significantly improving detection capabilities for both equipped and non-equipped vehicles and VRUs. Their methodology involves analyzing crash statistics from Japan, Germany, and the USA, identifying crash scenarios inadequately addressed by existing ADAS. The findings suggest that integrating CAS with CPS can substantially increase the effectiveness of ADAS, particularly in protecting VRUs and addressing vehicle-to-vehicle crashes in non-line-of-sight situations, thereby enhancing overall road safety.

\section{Scenario Outline}
\label{sec:scenarioOutline}
\begin{figure}[!h]
    \centering
    \includegraphics[width=\linewidth]{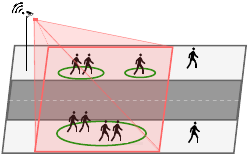}
    \caption{Schematic depiction of the scenario}
    \label{fig:schematicDepiction}
\end{figure}

In Figure \ref{fig:schematicDepiction} we show a schematic depiction of the scenario.
The dataset referenced in the preceding subsection supplies the movement data and projected future trajectories. 
It is assumed that VRUs are monitored by an RSU that is equipped with a camera. 
This camera captures data at an update rate determined by the dataset. Additionally, the RSU features connectivity capabilities. 
The formation of clusters is conducted offline using the algorithm described below.

For our study we will use the Dalian University of Technology (DUT) dataset \cite{yang2019top}, which comprises trajectories of vulnerable road users, specifically pedestrians, and vehicles.
This dataset was collected using a DJI Mavic Pro Drone equipped with a camera, positioned high above the target area to remain unnoticed by those below.
The video recordings are captured at a resolution of 4K with a frame rate of 23.98 frames/s.
Each trajectory entry in the dataset is accompanied by timestamped coordinates, along with supplementary attributes such as estimated speed, heading angles coordinates.

We employ a time-sequence DBSCAN algorithm\cite{hao2019} to identify clusters in the dataset.
At each timestamp, a cluster candidate is computed for a given VRU using DBSCAN \cite{martin1996}, and its constituent members are recorded.
A cluster is deemed valid only if the ratio of the clustering duration to the union of the durations of all its members exceeds a predefined threshold \textit{r}.
During the detection process, the algorithm iterates through all VRUs and identifies potential clusters.
Only those clusters whose members exhibit symmetric connectivity are considered as final clustering results. Here, symmetry means the group members are clustered no matter which of them is the given VRU. For example, within the iteration over users a, b, and c, clusters (a, b), (b, a), and (c, a) are identified. The clusters (a, b) remain fixed as they exhibit consistent clustering regardless of whether a or b is designated as the given VRU. Conversely, cluster (c, a) demonstrates connectivity solely when c is the designated VRU; hence, it is excluded from the final clustering results.
Consequently, the time-sequence DBSCAN requires two hyperparameters: the co-existence ratio \textit{r} and the Euclidean distance \textit{e}.
\begin{figure}[!t]
    \centering
    \includegraphics[trim=0.2cm 0.4cm 0.2cm 0.34cm, clip, width=\linewidth]{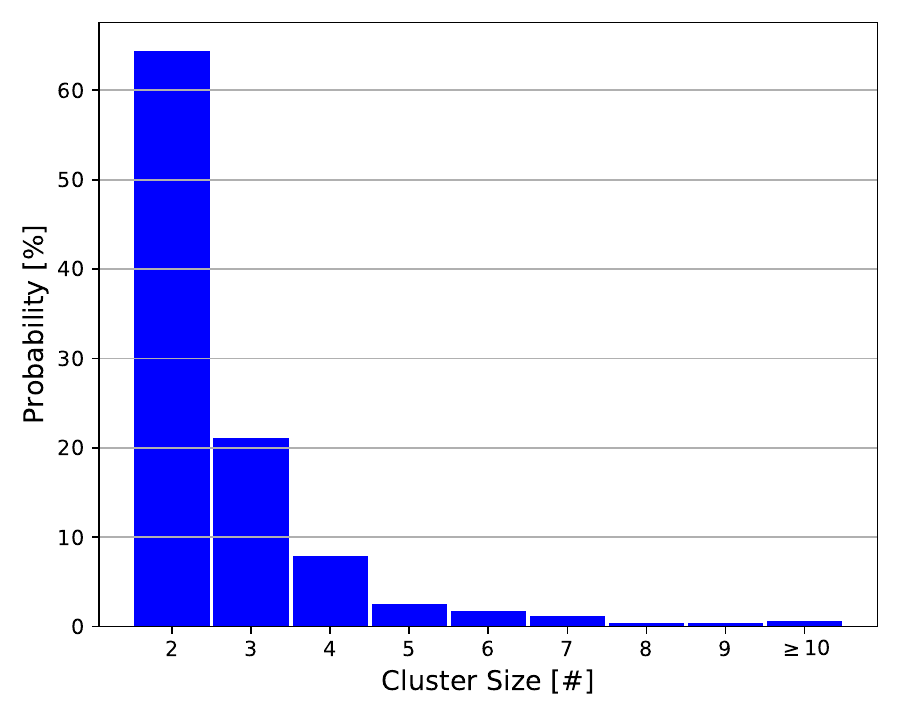}
    \caption{Probability Density Function for Cluster sizes in the dataset.}
    \label{fig:clusterSizeDistribution}
\end{figure}
For the DUT dataset, these hyperparameters are tuned by human experts based on the density and movement patterns of the VRUs.
For instance, reducing \textit{r} and increasing \textit{e} results in larger clusters, accommodating VRUs that are spatially distant and spend less time together while still being grouped.

In Figure \ref{fig:clusterSizeDistribution} we show the probability density function for cluster sizes.
For this statistic we count the clusters' sizes in each frame for the entire duration of the dataset.
The figure shows the size of the clusters formed by applying the algorithm described above.
It is noteworthy to mention that clusters with four participants or less count for more than 90\% of the clusters.
Individual VRUs are excluded from the probability density function calculation. However, our analysis reveals that in 27\% of instances, VRUs were not integrated into any cluster.

\section{Cluster Shapes Evaluation}
\label{sec:clusterShapes}
\begin{table}[!b]
  \centering
  \caption{Cluster shape size depending on included fields}
  \begin{tabular}{@{}lll@{}}
    \toprule
    \textbf{Shape} & \textbf{Full Option Included Size [B]} & \textbf{Compulsory Size [B]} \\
    \midrule
    Circle & 9 & 1.5 \\
    Ellipse & 12 & 3  \\
    Rectangle & 12 & 3 \\
    Polygon & 7.5 + NrPnts $\times$ 6 & NrPnts $\times$ 4 \\
    \bottomrule
  \end{tabular}
  \label{tab:clusterShapeSize}
\end{table}

\subsection{Cluster Shape in Context of  V2X}
\label{sub:clusterShapeV2X}
In the introductory section, we discussed how VAM and CPM effectively characterize entire groups of VRUs, representing these clusters through geometric shapes. Table \ref{tab:clusterShapeSize} outlines circular, elliptical, rectangular, and polygonal shapes as potential representations. V2X messages, formulated in ASN.1 format, also incorporate definitions of these geometric shapes. The table \ref{tab:clusterShapeSize} details the byte allocation required to depict different cluster shapes, both with and without optional fields. The definition of shapes utilizes specific fields, which may be optional. The size of shapes, excluding optional fields, is termed the compulsory size. Notably, the circular shape requires the minimum byte count, whereas the size of polygonal shapes varies with the number of points utilized for their description. For each type of cluster shape, the smallest shape that encompasses all participants is chosen. From a bird's-eye view, VRUs are presumed to be represented by a rectangular shape.
\begin{figure}[!t]
    \centering
    \includegraphics[trim=0.1cm 0.5cm 1cm 2.35cm, clip,width=\linewidth]{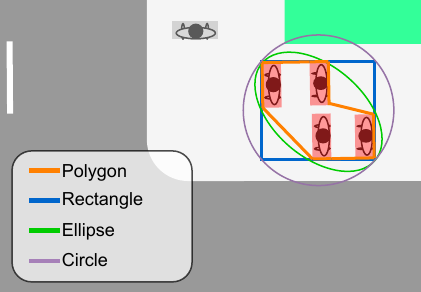}
    \caption{Bird-eye view of road section. Four VRUs have formed a cluster.}
    \label{fig:clusterShapeExample}
\end{figure}
For the bird-eye-view representation of persons we assume a fixed width of 50 cm and depth of 30 cm and retrieve the direction of VRUs from the dataset.
Different algorithms are employed based on the shape.
The Welzl algorithm is utilized for circular shapes \cite{welzl2005smallest}.
For ellipses, the algorithm from \cite{khachiyan1979polynomial} is chosen.
Convex hull algorithms, as summarized in \cite{preparata2012computational}, are used for polygonal and rectangular shapes.
In Figure \ref{fig:clusterShapeExample}, we present an example showcasing a cluster of four pedestrians, illustrating a potential scenario from the dataset that includes individuals. The bounding boxes around the pedestrians are depicted as transparent rectangles. Additionally, the four possible shapes used in this example are outlined around the pedestrians. It is readily apparent that the polygon shape occupies the least surface area among the shapes. However, the bit size required to represent the polygon depends on the number of points, indicating that its surface efficiency may incur a cost. Rectangles and ellipses exhibit similar characteristics, whereas the circle necessitates the largest surface area, despite being relatively efficient in terms of bit usage, as demonstrated in Table \ref{tab:clusterShapeSize}.

\subsection{Shapes Evaluation Metrics}
\label{shapesEvalMetric}
In this section we will introduce to the metrics used for the cluster shape evaluation.
The Clustering Accuracy (CA) metric quantifies the effectiveness of clustering shape description by measuring the proportion of VRUs correctly assigned to their respective clusters.
In the mathematical description, if it is correctly clustered, the i-th VRU accuracy score ($a_{i}$) is assigned a value of 1; otherwise, it is assigned a value of 0.
CA is then calculated by summing up all the accuracy scores for each VRU and dividing by the total number of VRUs which are under the shape N (Equation \ref{eq:CA}).
This yields a value between 0 and 1, where a higher CA value indicates a better clustering shape description with more VRUs correctly assigned to their respective clusters.
Intuitively it can be said that the metric provides insight into the accuracy of shape clustering description by quantifying the proportion of correctly assigned VRUs, thus aiding in the evaluation and comparison of different shapes.
\label{sub:shapesEval}
\begin{figure}[!b]
    \centering
    \includegraphics[trim=0.4cm 0cm 1.5cm 1.4cm, clip, width=\linewidth]{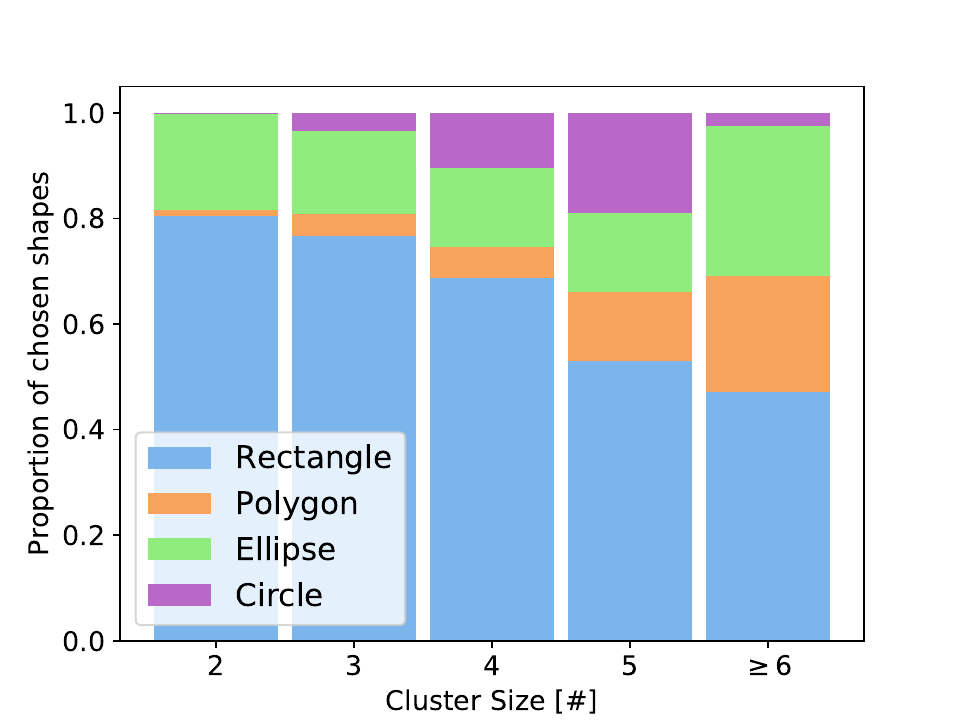}
    \caption{Shapes chosen by the adaptive algorithm depending on cluster size}
    \label{fig:shapeAssign}
\end{figure}
\begin{equation}
    \label{eq:CA}
    \begin{aligned}
         CA &= \frac{\sum_{i=1}^{N} a_i}{N}, \\
        \text{where } a_i &= 
        \begin{cases} 
            1 & \text{if the } i^{\text{th}} \text{ VRU is correctly assigned,} \\
            0 & \text{otherwise.}
        \end{cases}
    \end{aligned}
\end{equation}
In assessing the efficiency of cluster shape descriptions, the Comprehensive Area Density Information (CADI) metric has been employed.
It offers a quantitative measure of the effectiveness with which a cluster shape representation captures the spatial distribution of VRUs, while also minimizing the informational overhead.
Specifically, CADI combines the spatial density of VRUs within a cluster with the amount of information required to describe the cluster's shape. A lower CADI value signifies a more efficient description, indicating that the shape representation accurately reflects the VRUs' spatial distribution with minimal information. This metric allows for the comparison of different geometric representation of cluster shapes, highlighting those that best balance accuracy and information efficiency. The formula for CADI is given by:

\begin{equation}
CADI = \frac{S_{\text{{[bit]}}}}{N_{\text{{[VRUs]}}} / A_{[\text{m}^2]}} = \frac{S_{\text{{[bit]}}} \cdot A_{[\text{m}^2]}}{N_{\text{{[VRUs]}}}}
\label{eq:cadi}
\end{equation}
where \(S_{\text{bit}}\) denotes the size of the shape description in bits, \(N_{\text{VRUs}}\) is the number of VRUs within the cluster, and \(A_{\text{m}^2}\) represents the area of the cluster in square meters.

\subsection{Shapes Evaluation Results}

In the previous subsection, we introduced two metrics that can be used to quantify the accuracy and efficiency of the shapes describing a cluster.
We present the evaluation results for four different shapes.
Additionally, we introduce an adaptive algorithm. This algorithm operates by selecting different shapes to describe the cluster based on their accuracy and efficiency.
The operation of the adaptive algorithm is straightforward. Initially, it identifies the possible shapes, such as circle, rectangle, polygon, and ellipse.
Among these, it selects the shapes with the highest accuracy. Then, from the remaining options, it chooses the shape with the highest CADI value, indicating the highest efficiency.
The adaptive algorithm's choice of shapes is illustrated in Figure \ref{fig:shapeAssign}.
The results indicate a preference for the rectangular shape in describing clusters, particularly when these clusters are small.
This preference arises because smaller clusters are more likely to exhibit equal accuracy across different shapes.
\begin{figure}[!t]
    \centering
    \includegraphics[trim=0.45cm 0.5cm 0.48cm 0.35cm, clip, width=\linewidth]{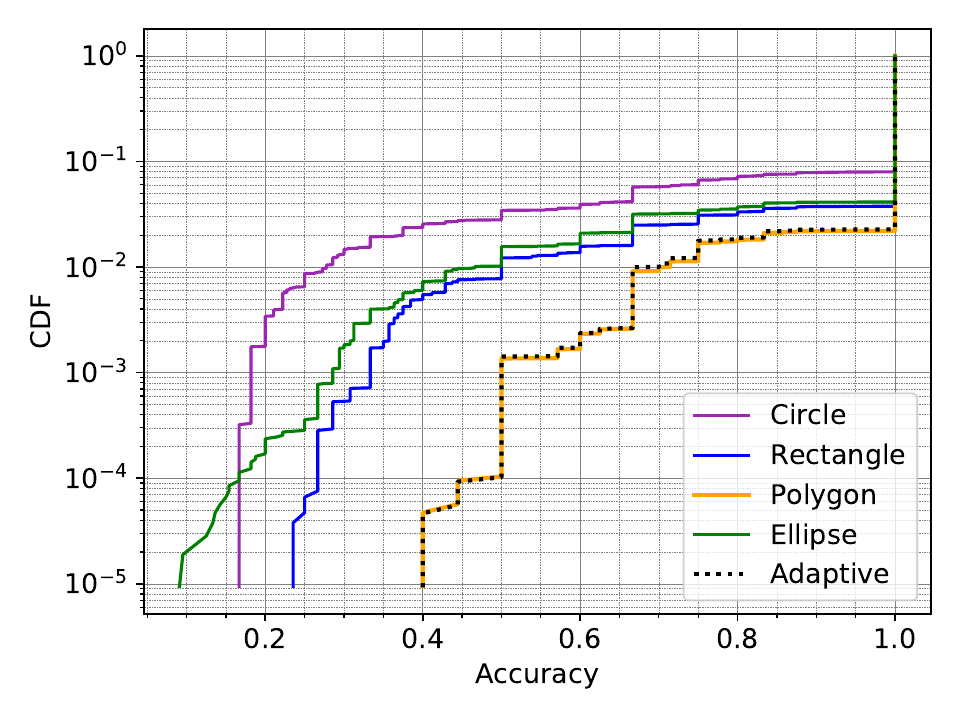}
    \caption{Shape accuracy description}
    \label{fig:accuracy}
\end{figure}
Consequently, the algorithm selects the shape with the highest CADI value.
Since a rectangle requires fewer bits for description compared to other shapes, it is chosen for the majority of cases.
As cluster sizes increase, the frequency of choosing rectangular shapes diminishes.
For polygon shapes, there is a noticeable steady increase in their selection as the cluster size grows.
This trend is attributed to the polygon shape's potential for higher accuracy, owing to its small surface area compared to the other shapes.
On the other hand, circular and elliptical shapes maintain a nearly constant cumulative usage across various cluster sizes.

In Figure \ref{fig:accuracy} we show the shapes' accuracy Cumulative Density Function.
CA is calculated as in Equation \ref{eq:CA}.
As expected, the accuracy for the polygon shape is the highest, since it also has the shape with the lowest surface compared with other shapes.
The smaller the surface needed to describe a cluster is, the higher are the chances for a better accuracy.
The adaptive approach has the same value for the accuracy, since it will always try to maximize its accuracy before choosing to increase the the information efficiency.
As for the other shapes, rectangle and ellipse show a very similar accuracy, which is also expected given that they have similar surfaces for small clusters.
The slightly higher surface of ellipse shapes compared to the rectangles does also reflect in the accuracy results.
Since circular shape does use more surface for describing a cluster, this is also reflected in the results.

\begin{figure}[!t]
    \centering
    \includegraphics[trim=0.45cm 0.5cm 0.48cm 0.35cm, clip, width=\linewidth]{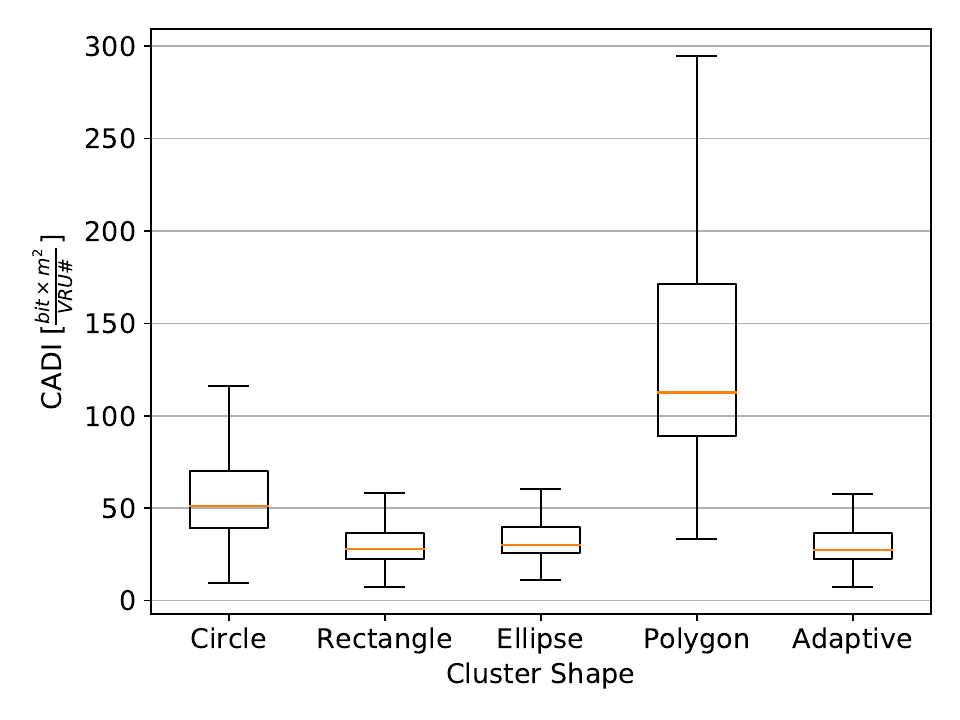}
    \caption{Comprehensive Area Density Information (CADI)}
    \label{fig:cadi}
\end{figure}
In Figure \ref{fig:cadi} we show the values for CADI, as calculated in Equation \ref{eq:cadi}.
As previously discussed, this metric serves as an approximate indicator of spatial and bit efficiency, reflecting the effectiveness of space utilization in conjunction with the quantity of transmitted bits. 
The rectangle and ellipse shapes demonstrate superior efficiency relative to the circle and polygon, with rectangles exhibiting the greatest efficiency. 
This superior performance of rectangles can be attributed to their high spatial efficiency, particularly in smaller clusters, which predominate in our dataset. Conversely, circles, despite requiring fewer bits, are significantly less space-efficient.

The polygon shape presents a unique case, possessing the highest CADI values and thus being the least efficient among all shapes. 
This inefficiency stems from the fact that the polygon's size in bits varies with the number of points describing it. 
Lastly, we want to emphasize how the adaptive shape maintains relatively high efficiency while achieving the highest accuracy, as shown in Figure \ref{fig:accuracy}. This outcome arises because the adaptive shape opts for the most efficient form unless a decrease in accuracy was previously observed.

\begin{table}[!b]
    \caption{Content of Collective Perception Message}
    \label{tab:cpmContent}
    \dirtree{%
        .1 \normalsize Collective Perception Message.
        .2 \small Message Header.
        .2 \small Management Container.
        .2 \small Originating RSU Container.
        .2 \small Sensor Information Container.
        .2 \small Perceived Object Container.
        .3 \footnotesize Perceived Object 1.
        .5 \scriptsize Class: VRU, Object Id.
        .5 \scriptsize Position: X, Y .
        .5 \scriptsize Velocity: X, Y Components .
        .5 \scriptsize Angles: X, Y, Z Components .
        .5 \scriptsize Dimensions: X, Y, Z .
        .5 \scriptsize Measurement Delta Time .
        .3 \footnotesize Perceived Object 2.
        .5 \scriptsize Class: VRU Cluster, Object Id.
        .5 \scriptsize Position: X, Y.
        .5 \scriptsize Velocity: X, Y Components.
        .5 \scriptsize Angles: X, Y, Z Components.
        .5 \scriptsize Cluster Shape.
        .6 \scriptsize Shape Dimensions.
        .6 \scriptsize Cardinality.
        .5 \scriptsize Measurement Delta Time.
        .3 \footnotesize \ldots \space Perceived Object N .
    }
\end{table}
\section{Collective Perception and Clustering}
\label{sec:results}
In the previous section, we evaluated the shapes used to describe clusters, focusing on their accuracy and efficiency. This section will adopt a more comprehensive perspective on the scenario, specifically examining the impact of clustering on the volume of data transmitted by an RSU. The scenario for this study is described in Section \ref{sec:scenarioOutline}. It is assumed that the RSU's placement allows for the detection of all VRUs as per the dataset.
Furthermore, it is assumed that the RSU possesses connectivity capabilities and is responsible for sending ETSI CPMs. The content of these CPMs is outlined in Table \ref{tab:cpmContent}. Notably, the CPMs will contain all standard containers, including the \textit{Originating RSU Container}, which is specific to messages sent by RSUs rather than vehicles. The \textit{Perceived Object Container} warrants particular attention; it includes information about the objects detected by the RSU. This container can list up to 255 objects of various classes, such as vehicles, VRUs, or clusters.
For the purposes of this study, the size of the CPM is considered excluding any optional fields, in alignment with the findings presented in \cite{xhoxhi2023first}. The shapes representing clusters are as specified in Table \ref{tab:clusterShapeSize}. According to the ETSI standard \cite{2etsiCPM}, the detection of a new VRU necessitates the generation of a new message encompassing all detected VRUs. Additionaly, VRUs should be included in a CPM message for at least every 500 ms. In our analysis, VRUs are included with a fixed frequency of 2 Hz.

\begin{figure}[!t]
    \centering
    \includegraphics[trim=0.45cm 0.5cm 0.48cm 0.35cm, clip, width=\linewidth]{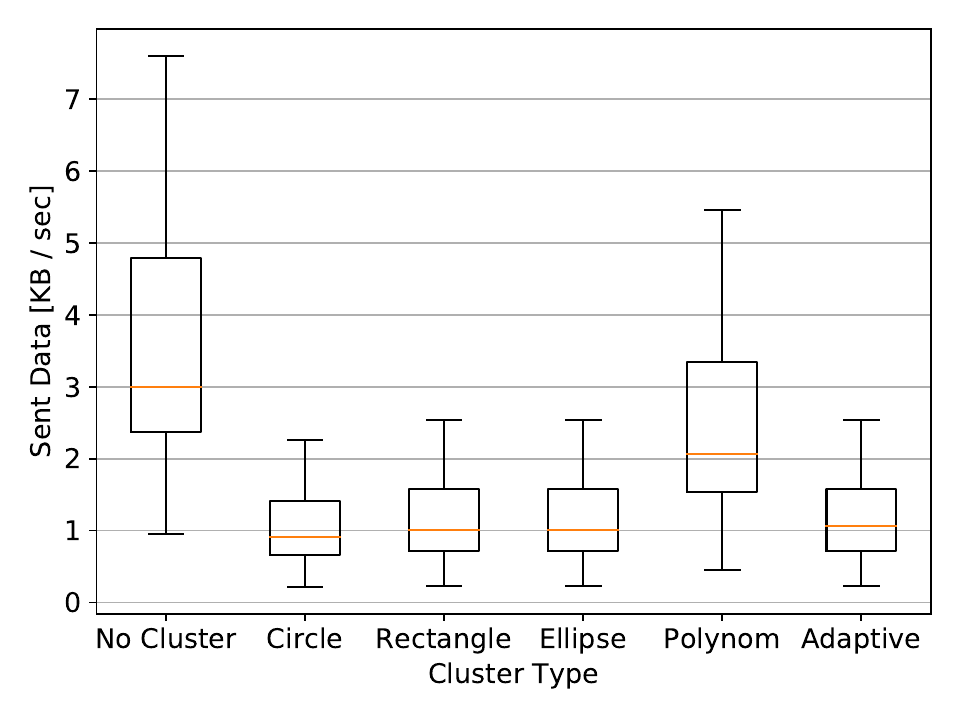}
    \caption{Data sent by the RSU in bytes/s depending on the cluster shape}
    \label{fig:dataSent}
\end{figure}

In Figure \ref{fig:dataSent}, we present the data transmission rates per second by the RSU across six different configurations. The \textit{No Cluster} configuration indicates that no clustering has been applied, and the VRUs are transmitted as distinct entities by the RSU. For the other configurations, we explore the utilization of various shapes and the proposed adaptive algorithm. VRUs outside of clusters are also transmitted as individual entities in the Cooperative Perception Message (CPM).
One immediate observation from the results in Figure \ref{fig:dataSent} is the clear benefti of employing clusters, as the median data transmission rate can be reduced by up to two-thirds. Another noteworthy observation is the reversal of roles between circles and rectangles in their comparison between the CADI effectiveness and the data transmission rate (bytes/s). This further underscores that circles encompass a larger surface area compared to rectangles. The adaptive algorithm demonstrates a median data transmission rate that is one-third of that observed in the \textit{No Cluster} configuration.

\section{Conclusions and Future Outlook}

In this work, we have evaluated the impact of utilizing various geometric shapes to describe VRU clusters. We assessed the potential shapes in terms of accuracy and efficiency. To this end, we introduced the Cluster Accuracy (CA) and Comprehensive Area Density Information (CADI) metrics. A simulation study was conducted using a scenario derived from a real-world dataset. Our results indicate that the polygonal shape offers the highest accuracy in cluster description but exhibits the lowest efficiency, making it more suitable for describing larger clusters. Conversely, a rectangular shape demonstrates improved efficiency. Additionally, this study presents an adaptive algorithm that selects shapes based on their accuracy and efficiency, prioritizing accuracy. We conclude the study by illustrating the clustering impact on channel load, comparing the bytes of data transmitted each second. Our results show that in the specified scenario, the data sent per second can be reduced by up to 65\% in median values. Future work will explore extending our study to include the computational power required for calculating the sizes of different cluster description shapes. Also the tracking accuracy of VRUs dependent on the chosen shape is an open topic.

\section*{Acknowledgment} This publication was funded by the Deutsche Forschungsgemeinschaft (DFG, German Research Foundation) - project number 227198829 / GRK1931 and by the Lower Saxony Ministry of Science and Culture under grant number ZN3493 within the Lower Saxony “Vorab“ of the Volkswagen Foundation and supported by the Center for Digital Innovations (ZDIN).

\bibliographystyle{ieeetr}
\bibliography{bibliography}

\end{document}